# A Large-Scale Analysis of IoT Firmware Version Distribution in the Wild

*Short Paper, Work in Progress*

*Frank Ebbers*

*Fraunhofer Institute for Systems and Innovation Research ISI,*
*Breslauer Str. 48, 76139 Karlsruhe, Germany*
*frank.ebbers@isi.fraunhofer.de*

This paper examines the up-to-dateness of installed firmware versions of IoT devices accessible via public internet. It analyzes datasets of 1.06m devices collected from the IoT search engine *Censys* and maps the results against the latest firmware version each manufacturer offers. By applying the SEMMA data mining process, a fully scalable and adaptive approach is developed. This approach relies on three data artifacts: raw data from *Censys*, a mapping table with firmware versions and a keyword search list. The preliminary results confirm the heterogeneity of connected IoT devices. They show that manufacturer, device type and country influence the up-to-dateness of firmware. The results suggest users as a "weak link" as they do not update the firmware of their devices in a timely manner. However, the heterogeneity leads to results not showing a high reliability, yet.

***Keywords****: Internet of Things, IoT, firmware, version, patch, update, up-to-dateness*

## Introduction

As of 2020, Internet of Things (IoT) devices can be found in almost every walk of life. Notwithstanding the benefits these devices offer to society and economy, they come with a plethora of security challenges. As most manufacturers prioritize rapid development over comprehensive security, the device firmware (FW) tends to be plagued with vulnerabilities. These can be exploited by criminals, for example to hijack the device for botnets and distributed denial-of-service (DDoS) attacks with ever increasing power (Antonakakis et al. 2017).

The most important means to fix FW vulnerabilities are updates and patches (U&P). While software patching is quite common practice for computers and mobile devices, FW updates for IoT devices are much less common, since they are comparatively hard to carry out (Antonakakis et al. 2017; Yu et al. 2015). For this reason, users tend to avoid the installation of firmware patches even if they are available (Vaniea and Rashidi 2016). Detailed studies find that only very few users perform firmware updates (e.g. Neshenko et al. 2019). To make matters worse, many manufactures not to provide patches in timely manner (e.g. Shahzad et al. 2019). This highly unsatisfactory state of affairs has since raised the attention of policy makers. In 2019, the European Parliament passed a consumer protection directive entitling IoT users (amongst others) to receive updates within a reasonable timeframe (*Directive (EU) 2019/770*, 2019).

In order to improve IoT security across the board, it is crucial to first detect and identify IoT devices with outdated firmware. This, however, is not a trivial process because no uniform patterns or methodologies exist (Antonakakis et al. 2017). Despite these difficulties, several authors have attempted to examine the type of installed FW versions for a specific printer (Cui et al. 2013). Other researches have been able to identify a distinct open source FW for routers (Wang et al. 2019), but not investigating the version number.

To the best of our knowledge, no comprehensive study about FW distribution and up-to-dateness of IoT devices has been carried out so far. We address this gap by performing a large-scale analysis of firmware version share by using data from the search engines *Censys* (Censys 2020) and *Shodan* (Shodan 2020) that gather information about IoT devices "in the wild", that is, accessible through the Internet. Accordingly we ask the following research question: *How up-to-date are IoT devices in the wild with regard to their* *installed firmware version and is there a difference between countries and device types? Our study also addresses the question whether the users or manufacturers are the main culprits of deprecated installed firmware in order to identify the "weakest link" in the IoT.*

The remainder of this short paper is structured as follows: first we motivate the need for IoT firmware U&P and describe challenges for their deployment. We then explain our methodology and present our preliminary results, which are discussed in a subsequent chapter. We conclude our work with suggestions for improvement and outline further research.

## State of Research

### IoT Security Challenges and the Need for Updates and Patches

While security is considered a major challenge within the IoT, research in this area is still in its infancy (Neshenko et al. 2019). Various authors coincide that IoT devices are easy targets for remote attacks (e.g. Kolias et al. 2017; Ransbotham et al. 2016; Yu et al. 2015). Most of these devices have very limited hardware capabilities to resist security threats (Ray et al. 2017). This makes it easy for hackers to exploit known security vulnerabilities, intrude in even very sensitive areas (e.g. users' homes) and infect millions of connected devices at once to perform DDoS attacks (Antonakakis et al. 2017; Kolias et al. 2017).
Kolias et al. (2017, p. 83) identified five reasons why IoT devices are "particularly advantageous for creating botnets". 1) They are constantly online. 2) Usability is prioritized over security. 3) Collectively, IoT devices are powerful to create enough traffic and 4) due to the absence of control interfaces, attacks often go unnoticed. Lastly (5), devices are poorly updated and patched, turning zero-day into eternal vulnerabilities. The longevity of IoT devices exacerbates this effect. Typically, the firmware of IoT devices is not a well-tested product, so security vulnerabilities typically exist in deployed code (Ransbotham et al. 2016). To fix these, U&P are a common and important remedy (Vaniea and Rashidi 2016), since they tend to have a significant positive impact on IT system security. For example, a survey of 3,000 IT professionals worldwide found that 60 percent of data breaches could have been prevented by patching (ServiceNow 2018).

### Patching and Updating the Internet of Things

Patching the software of personal computers (PC) and smartphones is a well understood process and has been researched from the users' as well as the manufacturers' perspective (Antonakakis et al. 2017; Ray et al. 2017). For PCs, studies show that automatic updates highly increase installation rates (Vaniea and Rashidi 2016). However, over-the-air (OTA) updates are difficult to implement in IoT devices because of limited storage, processing power, and their high inter-connectivity (Ray et al. 2017; Yu et al. 2015), which vary between different IoT device types. The same applies for their interfaces and methods for user update notification (Cheng et al. 2017), which are difficult to implement. We therefore hypothesize that *there could be marked differences in the up-to-dateness of the firmware for diverse IoT device types (H1).*

The processes and frequency for distributing patches varies between vendors. E.g. Antonakakis et al. (2017) find that some manufactures assume a higher degree of responsibility for infected devices than others. Thus we hypothesize that *the installed latest firmware versions differs between manufacturers, regardless of the device type (H2).* Other research finds that manufacturers' "patch release behavior is an underinvestigated component of overall software quality and security" (Arora et al. 2010, p. 116). Several studies, e.g. by Shahzad et al. (2019) find that there are significant delays between vulnerability disclosure and patch release. By translating this observation to the field of IoT, we hypothesize that *there are IoT device manufacturers who do not provide regular updates or patches (e.g. in annular intervals) for devices that are still in use and online (H3).* Vaniea and Rashidi (2016) observe that users often hesitate to install U&P due to lack of motivation or technical illiterateness. A study of 2,000 IoT owners finds that around 40 percent never perform firmware updates and further 10 percent even don't know what firmware is (Canonical 2017). As a prior study shows that users delay software updates for 80 days (Vitale et al. 2017), we hypothesize that *even if firmware updates are available, many users do not install them within two months after the release and thereby constitute a weak link in the IoT security (H4).*

### Firmware and Software Version Share

Several studies examine the distribution of different versions of web browsers or mobile and PC operating systems (StatCounter 2020). A recent study finds that 55 percent of all installed PC programs worldwide are out-of-date (Avast 2019). Similar research efforts are made in order to identify the firmware and software versions of web servers by employing dedicated tools such as WhatWeb (WhatWeb 2019).

In the IoT field, the tool IoTTracker by Wang et al. (2019) scans for specific open-source firmware for routers, classified by vendor and country, but provides no insight about the installed firmware version. For a specific printer, Cui et al. (2013) find that only 1.08% them run the latest FW. However, to the best of our knowledge, there is no comprehensive study on the version share of IoT devices, yet.

According to Kumar et al. (2019, p. 1169) "[d]evice types and manufacturer popularity vary dramatically across regions". IT security behavior differs also between cultures (Chen and Zahedi 2016), as witnessed e.g. in how frequent users install updates on PCs (Vitale et al. 2017). Our last hypothesis therefore states that *there are regional differences in the up-to-dateness of installed firmware versions of IoT devices, which are independent from device type and manufacturer (H5).*

### IoT Device Discovery and Annotation

In order to protect IoT devices effectively, it is crucial to first identify their type, manufacturer and model name (Feng et al. 2018). This is e.g. exemplified by a 2019's study, which found that devices produced by same vendors or series usually come with similar security vulnerabilities (Wang et al. 2019). There are different approaches to discover such information, e.g. trough web crawling, natural language processing and data mining implemented by Feng et al. (2018) and Wang et al. (2019). Kumar et al. (2019, p. 1170) find that "90% of devices worldwide are produced by only 100 vendors".

There are also search engines that specialize in IoT devices. Two prominent ones are *Shodan* (Shodan 2020) and *Censys* (Censys 2020). Both scan IP addresses, different ports and protocols and grab banner data, as well as other meta data, e.g. device's location. *Shodan* was released in 2009 and scans the IPv4 and IPv6 space. Their web crawlers generate a random IP address and test a random port for accessibility. If successful, the banner information is grabbed. This strategy ensures the coverage to be relatively uniform. To limit traffic, increase search speed and avoid geographical bias, the crawlers are located in different countries (Matherly 2018). *Shodan's* database supports queries with pre-filters (e.g. constraining a search to a particular city). *Censys*, which was established in 2015 by a team of researchers from the University of Michigan, works similar to *Shodan*, but its scans are limited to the IPv4 address space. Unlike *Shodan*, it is open source and can be used freely for academic purposes. The infrastructure and storage is provided by Google. *Censys* pings four billion devices each day. The data is aggregated on a daily basis and accessible via Google's BigQuery data warehouse, allowing to perform historical searches that go back until 2018. Similar to *Shodan*, the datasets includes amongst others the device type, manufacturer and device location. *Censys* provides two different data tables; one aggregated data table "banner" and a detailed "public" table.

### Research Question and Hypotheses

As outlined above, our research question asks, h*ow up-to-date IoT devices in the wild are, with regard to their installed firmware version and if there is a difference between countries and device types?* It also asks *whether the users or manufacturers are the main culprits of deprecated installed firmware*.

Our literature review suggest that there is no comprehensive study on the frequency and up-to-dateness of firmware versions, updates and patches on IoT devices. We address this gap with this large-scale study on IoT devices and their respective firmware version. In view of the available literature and the theoretical discussion above, our working hypotheses are as follows:

H1: There could be marked differences in the up-to-dateness of the FW for diverse IoT device types.

H2: The installed base of latest FW versions differs between manufacturers, regardless of the device type.

H3: There are IoT device manufacturers who do not provide regular updates or patches (e.g. in annular intervals) for devices that are still in use and online.

H4: Even if FW updates are available, many users do not install them within two months after the release and thereby constitute a weak link in the IoT security.

H5: There are regional differences in the up-to-dateness of installed firmware versions of IoT devices, which are independent from device type and manufacturer.

# Methodology

Our data mining approach followed the sequence of steps from the SEMMA process developed by the SAS Institute. It has the advantage of being iterative, applicable across a variety of industries and of offering methodologies for diverse business problems (SAS Institute 2017). We employed the storage and services facilities of the Google Cloud and its services, in particular BigQuery. In summary, our analysis was based upon three data artifacts: real-world data from *Census*, a list of keywords and regular expressions to filter the datasets, and a table mapping between firmware versions and device models.

### Step 1: Sample Generation, Dataset and Selection of Search Terms

Although not specified in the SEMMA process, we first generated a data basis. We contacted top IoT manufacturers (e.g. Amazon, Google, HP, Siemens, and Bosch). However none were willing to provide information about the up-to-dateness of their devices. Thus we focused on the IoT search engines *Shodan* and *Censys*, to ensure non-biased and good-quality data. As both databases represent the current real IoT distribution in the wild, this is advantageous for our analysis.

In order to create a list of suitable search terms that reveal information about a firmware's version, we applied an iterative process. First, we downloaded the results of a simple search with the term "firmware version" from *Shodan*, which resulted in 17k device datasets. This dataset also includes the source code of the HTML devices' administration interface or telnet login screen. We conducted a keyword analysis of the source codes, utilizing the Top-K Sequential Patterns algorithm by Fournier-Viger et al. (2013). This yielded multiple keywords and HTML-tag-names carrying information about installed firmware versions (e.g. class="fw-version"). In addition, we used *Shodan* Facet Analysis (Shodan.io 2020) to identify keywords. We also determined IoT manufacturers and device models known to have been hijacked by botnets in the past (Antonakakis et al. 2017; Guo and Heidemann 2018). These were added to the list of search keywords. As a last step, we translated these into regular expressions (regexp), if applicable.

We selected regexp that allowed to directly draw conclusions about the installed firmware. We selected regular expressions that allowed to infer the installed FW version (e.g. "Firmware ver. /d*./d*"). We then ran database searches in February and mid of April 2020 on the public and banner tables of *Censys*, which resulted in 1,06m devices. This sample constitutes the base dataset used in the processing steps described below. Duplicate devices were removed. We calculated the statistically representative sample size for this N=1,06m, using standard sample calculation: 16,384 devices at a confidence level of 0.99, with confidence interval=0.1 and crated it using the RAND-function in Google BigQuery.

### Step 2: Exploring - Initial Characterization of the Sample Data Set

In view of our working hypotheses, we first analyzed the manufacturer and device type fields. We found a small number of manufactures to be strongly overrepresented (e.g. MikroTik, ZTE and D-Link). It also turned out that the majority of the devices were internet routers. Some five percent of the sample values did not contain any information about device type, product or manufacturer at all. The data fields "revision" and "version" did not contain any suitable information. Most devices were installed in Venezuela (6.3%), Brazil (4.3%) and Russia (4.1%). Nevertheless, this country-field contained null-values in 42% of all cases.

### Step 3: Modify - Adjustments to the Sample Data Set

As a next step, we introduced a number of modifications on data fields. We first created a unique identifier for each dataset by calculating a cumulative hash value across all fields. We then focused on missing field values in *country*, *manufacturer*, *device model* and *type*. E.g., we decided to use the *country-code* field instead of *country*, as it showed only 7.1% null-values. Next, we scanned the source code for manufacture names and models, since this approach had been successfully applied by Kumar et al. (2019). To amend missing values on manufacturer, device model and type, we asked Wang et al. (2019) to classify our devices, which resulted in 217,836 successfully classified devices.

Our search strategy employed very precise regular expressions and yielded the base dataset (1.06m). Applying the Top-K Sequential Patterns algorithm again provided additional patterns in the devices' source code that allowed to further update and extend the search keyword list to also include the identification of device models. In order to extract the firmware version, we first we used the regular expressions developed for the database queries to extract strings containing firmware information from the full base dataset. Then we extracted patterns resembling firmware version numbers. In a similar fashion, we developed patterns representing firmware deployment dates in the source code. In order to create a mapping table containing the device models with their current, as well as former firmware versions, and their respective deployment dates, manufacturers' websites were browsed manually to extract the necessary information.

### Step 4: Modelling - Mapping Devices with Firmware Dates

The final artifact of our SEMMA process is a table containing device type, model, manufacturer, general FW date without version information (gFD), installed FW Version (iFV), installed FW date (iFD), latest FW version (lFV), latest FW date (lFD), age installed FW as of April 2020 (AiF), age latest FW as of April 2020 (AlF) and time between iFW and lFW (TB). Not every device provided all information, so we applied the rules presented in Table 1. We conducted a variance analysis of the AiF and AlF for our independent variables (country, device type and manufacturer) and calculated the effect size for each of them.

| Available: | | | | | Rule: |
|---|---|---|---|---|---|
| gFD | iFV | iFD | lFV | lFD | * Rule is not processed yet. |
| | x | x | x | x | lFD - iFD = TB [months], if lFD = iFD: firmware is up-to-date |
| | | x | | | Today - iFD = AiF [months] |
| | | | x | x | Today - lFD = AlF [months] |
| x | | | | | Today - gFD = AiF [months]* |
| | x | | x | | Comparing the FW version number: Major (3.0→4.0) or minor version jump (3.0→3.1)* |
| | x | | | | No calculation possible |

**Table 1. Modelling rules**

### Step 5: Assessing - Model Reliability

Usually a model's reliability is tested by applying it to a different set of data. However, due to time constraints, we were not able to provide reliability measures at this point of time.

## Results

In total, 1,061,284 devices fit our regular expressions. This include 113 distinct models (e.g. "DIR-860L") from 63 distinct manufacturers, which can be assigned to 14 device types. It should be noted that many devices are routers, hence, not IoT devices in the strict sense of the term. Most devices in this list are installed in Venezuela, Brazil and Russia. No country information is provided for 7% of the devices from the list, while 26% do not provide the device type. The mapping table consists of 1,899 devices' firmware versions or dates for 401 distinct devices, of which 296 contain FW versions and dates. There are more distinct models in the mapping dataset, because it includes all available models of a device's series already.

Our approach extracted a version or date for 1,055,083 of all devices (99.4%), with a majority of devices from MikroTik (57%) and ZTE (20%). We matched the installed version and date for 460,773 devices (42%) of 12 device categories, showing a majority for network devices (53%). The overwhelming majority of these devices were produced by MikroTik (99%). 226,785 (49%) of all run the latest available firmware version. The **up-to-dateness (age) of the installed firmware (AiF)** represents the number of months between the deployment date of the installed version and April 2020. The results show that the average age of the installed FW version is 19.2 months. Smart Home devices (77.0 months) show the highest average age, whereas access points are relatively up-to-date (11.0 months). Table 2 shows that devices from Hikvision show the highest average age (96 months), whereas Algo and Cisco devices are relatively up-to-date.

Considering differences between countries, devices in the Bahamas are relatively rarely updated (avg. 64.38 months), whereas Haitians and Liechtensteiners update their devices on average every second months.

| Device type | Ø age | Manufacturer | Ø age | Country | Ø age |
|---|---|---|---|---|---|
| smart home | 77.00 | Hikvision | 96.00 | Bahamas | 64.38 |
| misc | 70.00 | Dahua Technology | 80.00 | Ethiopia | 55.20 |
| ... | | ... | | ... | |
| industrial controller | 13.58 | Cisco | 11.0 | Liechtensteiners | 2.00 |
| access point | 11.00 | Algo | 11.0 | Haitians | 2.00 |

**Table 2. Top and last two average age of installed FW in months (by device type, manufacturer and country), overall avg = 19.2**

Descriptive statistics and a univariate ANOVA show the effect sizes and significance visualized in Figure 1. However it is likely that effects are not estimated correctly, as the number of devices within groups varies dramatically (e.g. 57% are MikroTik devices, whereas other 32 manufacturers provide less than 0.01% of all devices).

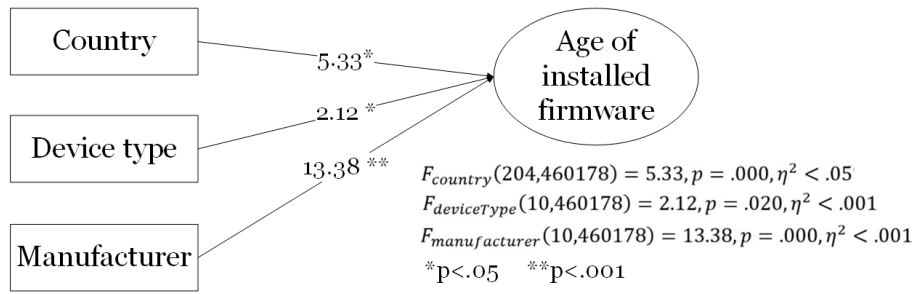

$$F_{country}(204,460178) = 5.33, p = .000, \eta^2 < .05$$
$$F_{deviceType}(10,460178) = 2.12, p = .020, \eta^2 < .001$$
$$F_{manufacturer}(10,460178) = 13.38, p = .000, \eta^2 < .001$$

*p<.05    **p<.001

**Figure 1. Results of univariate ANOVA**

Reading the **up-to-dateness (age) of the latest firmware (AlF)**, the results show that manufacturers Brother (43.5 months) and HP (37.2 months) show the highest average age of all devices, whereas MikroTik's firmware are very frequently updated (<1 month). Smart home devices' firmware are oldest (60.3 months), whereas network devices are kept most up-to-date by the manufacturers (0.004 months). The **time between installed and latest firmware (TB)** varies between 0 and 102 months with an average time of 18.88 months (σ=21.32).

## Discussion and Contribution

Our preliminary results suggest that connected IoT devices are diverse in type, manufacturer and country. If tailored regular expressions are applied, the firmware version identification rate can be brought up to 99.4%, notwithstanding that there is currently no uniform way of gathering FW information (Antonakakis et al. 2017). Identifying the correct FW for a device is rather complicated, though, as there are different FW versions for various device revisions and models sold in specific countries.

The wide range of maturity of the installed firmware versions among device types (σ=26.83) suggests that some devices provide much better update and patching mechanisms than others, as stipulated in H1. This could be attributed to a lack of interfaces or a "setup-and-forget" mentality (Cheng et al. 2017; Kolias et al. 2017). The results show that the average age of FW varies between countries (σ=8.65) and proves that system security is perceived differently between countries, which supports our hypothesis H5. Similarly, the average ages of the latest available FW version differs strongly between manufacturers (σ=22.60). Notably, the manufacturer size appears to be insignificant. This is in accordance with results from Arora et al. (2010) and supports our hypothesis H2. Some manufactures may be less eager to patch or users run many devices of specific manufacturers that already reached end-of-support. As our data does not indicate these devices, we cannot fully test the validity of H3. The time between availability and installation of firmware updates (TB) is an important indicator for user acceptance. Our results suggest that, on average, a firmware update is installed every 18.6 months, independent of device type and manufacturer. As this is significantly more than our threshold of two months, these results confirm our hypothesis H4. Users can therefore be considered as a weak link in IoT security. As a caveat, we note that because of the dominance of same values, the results are prone to misinterpretation, e.g. there are many devices from MikroTik, whose avg TB (18.9) is already close to the overall TB. Surprisingly, 49% of the devices run the latest available firmware version. However, we found that it is mainly driven by the router "H108N V2.5", whose version number corresponds not to the firmware version, but rather than the edition. Excluding this device, the up-to-dateness rate drops dramatically to 2.45%, which seems to be more realistic.

Our contribution can be applied in practice and theory. For example, it suggests to develop methods for nudging users to install latest FW. Practitioners could apply our approach in regular intervals to identify trends, e.g. after FW update regulations were passed or to benchmark firmware distribution methods.

Some limitations of our current approach should be highlighted. First, the reliability of our results is still limited due to overassessment. This problem will be addressed in the final paper, while this report just describes work-in-progress. Second, the majority of devices analyzed so far are not IoT devices in the narrow sense of the word. In order to address this problem, our process has to be extended with methods of IoT fingerprinting and more precise search expressions. Lastly, we do not assess the impact of patching vulnerabilities for specific firmware versions. To address this issue, the firmware we identified would have to be matched against common vulnerability databases.

## Conclusion and Outlook

Identifying IoT devices firmware is a step towards overall IoT security. To the best of our knowledge, we have carried out the first comprehensive study about FW distribution and firmware up-to-dateness of IoT devices. We presented an iterative and scalable approach to identify IoT devices' firmware versions from *Censys* data. Our preliminary results suggest that the up-to-dateness of IoT firmware is influenced by device types, manufacturers and country of installation. While our current results are still too tentative for providing reliable effect sizes, we are optimistic that they will become more precise once more devices are analyzed. In summary, we found empirical support for our hypotheses H1, H2, H4, and H5, while H3 remains unanswered due to lack of suitable data.

Future work could include information from vulnerability databases to analyze the severity of vulnerabilities in old firmware versions. Further, we encourage researchers to apply and refine our data-mining approach. This could shed light on the general trend of firmware updates or to measure the effect of regulations.

## Work to be finalized

We will extend our database by integrating the results from the *Shodan* database. We will refine our list of search terms in order to identify additional devices. More device models will be added to the mapping table to provide more precise results. We are optimistic, that this will help to provide more reliable measured values. All this will help to picture the heterogeneous IoT market. As our approach is fully scalable, the analysis will not result in more programming work.

## Acknowledgements

This work is funded under the SPARTA project, which has received funding from the European Union's Horizon 2020 research and innovation programme under grant agreement No. 830892.